\documentclass{article}
\usepackage{amssymb}
\newcommand{\be}{\begin{equation}}
\newcommand{\ee}{\end{equation}}
\newcommand {\ba}[1]{\left( \begin{array}{#1} }
\newcommand {\ea} {\end{array}\right )}

\begin{document}

\title{{\bf Polarization Elements--A Group Theoretical Study}}
\author{{\bf  Sudha and  A.V.Gopala Rao} \\
\\
Department of Studies in 
Physics \\ 
University of Mysore \\
Manasagangothri \\
Mysore 570 006}
\date{}
\maketitle
\begin{abstract} 
The classification of 
 polarization elements, the polarization affecting optical devices
 which have a Jones-matrix representation, according to the 
 type of eigenvectors they possess, is given a new visit through 
 the Group-Theoretical connection of polarization elements. 
 The diattenuators and retarders are recognized as the elements 
 corresponding to boosts and rotations, respectively.
 The structure of homogeneous elements other than diattenuators 
 and retarders are identified by giving the quaternion corresponding
 to these elements.
 The set of degenerate polarization elements is identified with the 
 so called `null' elements of the Lorentz Group. 
 Singular polarization
 elements are examined in their more illustrative 
 Mueller matrix representation and finally the eigenstructure of 
 a special class of singular Mueller matrices is studied. 
\end{abstract} 

\noindent{\bf Key words:} {\em Homogeneous, Inhomogeneous, Degenerate polarization elements, 
Eigenpolarization,\\ 
Lorentz Group, Quaternions, 
Singular Mueller matrices}

\section{Introduction}
It is well known that \cite{S, LC} polarization elements are 
characterized by the types of eigenpolarization that they possess.
\emph{Homogeneous polarization elements} are the ones which possess 
orthogonal eigenpolarization whereas \emph{inhomogeneous polarization 
elements} possess non-orthogonal eigenpolarizations.
Here we refer to a \emph{polarization element} as a 
polarizing optical device which has got the Jones matrix or the 
$2\times 2$ matrix representation. Also, the term eigenpolarization 
refers to the eigenvectors of the associated
Jones matrix, the states which are unchanged in polarization by 
the action of the corresponding Jones device matrix. 
There is another class of polarization
elements that are called \emph{degenerate} polarization elements.
They are the ones which possess only one linearly 
independent eigenpolarization.

Though the connection between the Lorentz group and 
polarization elements is well known \cite{B, Mee, SS, CL}, little 
has been done in exploiting the known properties of
the Lorentz group in identifying the homogeneous, inhomogeneous and 
degenerate elements. Ours is an attempt towards this end and by 
achieving this, we hope to have lessened the jargon to those 
who are familiar with Group Theory.

We recall here the connection between the set of all 
non-singular polarization elements
and the Lorentz group \cite{Mee}. 

\noindent \textbf{A non-singular pure Mueller matrix is of the 
 form $k{\bf L}$ where $\bf L$ is an element belonging to the 
 Orthochronous Proper Lorentz Group (OPLG) SO(3,1), 
 with $k$ being any positive real number.} 

\noindent Though the connection is always stated in terms of 
 $4\times 4$  representation of the polarizing optical devices, 
 the so called Mueller matrix representation, and the group SO(3,1), 
 the connection between the $2\times 2$ matrix representation of 
 the device (Jones representation) 
and the group SL(2,C) is obvious through the homomorphism between 
the groups SL(2,C) and SO(3,1). In fact, there is a one-to-many mapping 
between the set of all  SL(2,C) matrices and the set of all Jones 
matrices, a Jones matrix being just a complex scalar times a SL(2,C) 
matrix. This being the case, a classification of SL(2,C) into 
homogeneous and inhomogeneous
elements gives us the corresponding classification of 
non-singular Jones matrices.
\section{Classification of elements of 
SL(2,C) on the basis of their eigenvectors }

We know that among the elements of SL(2,C), we have those 
elements which are unitary and hence correspond to the subgroup 
SU(2) of SL(2,C).
The elements of SU(2) written in their quaternionic \cite{KNS} 
representation\footnote{The quaternions are mathematical objects 
of the form ${\bf q}=q_1{\bf e}_1+q_2{\bf e}_2+q_3{\bf e}_3+q_0\equiv 
\vec{\bf q}{\bf e}+q_0$ where  ${\bf e}_1$, $\bf {e}_2$  and 
${\bf e}_3$  are symbols called the 
\emph{quaternion units} satisfying the relations 
${\bf e}_i^2=-1,\; (i=1,\;2,\;3)$, ${\bf e}_i {\bf e}_j={\bf e}_k 
(i\neq j\neq k$ and $i,\;j,\;k$  cyclic. The quaternions over the 
field of complex numbers are called complex quaternions and they 
form a group under multiplication. A detailed discussion on 
quaternions and how they form a representation of the Lorentz group 
can be found in \cite{KNS}} are given by
\begin{equation}  
{\bf{T}}_{r}=\ba{rr} q_{0}-iq_{3} & -iq_{1}-q_{2} \\
       -iq_{1}+q_{2} & q_{0}+iq_{3} \ea
\end{equation}
 where 
$q_{0}$, $q_{1}$, $q_{2}$ and 
$q_{3}$ are all real and ${\bf q}\cdot {\bf
q}=q_0^2+q_1^2+q_2^2+q_3^2=1$. 
It is very easy to notice that 
$\bf {T}_{r}$ has got two eigenvalues 
$ \lambda=q_{0}\pm \sqrt{q_{0}^{2}-1}$ and owing to the relations 
\cite{KNS} 
$q_{0}=\cos {\theta\over 2}$,
$q_{i}=a_{i} \sin{\theta\over 2}$ ($i=1,\;2,\;3$),  
$\hat {\bf a}\cdot \hat {\bf a}=1$, where 
$\hat{\bf{a}}=\{a_{1},\;a_{2},\;a_{3}\}$ is a real (unit) vector,    
we have 
$\lambda_{1,\;2}=\exp {\pm{i\theta\over 2}}$.
The eigenvectors of $\bf{T}_{r}$
belonging to these eigenvalues are found to be 
\begin{equation}
{\bf{X}}_{1}=\ba{c} 1 \\ 
\frac{a_{1}+ia_{2}}{a_{3}-1} \ea;\quad  
{\bf{X}}_{2}=\ba{c} 1 \\ 
\frac {a_{1}+ia_{2}}{a_{3}+1} \ea
\end{equation}
which satisfy $\bf{X}_{1}^{\dagger}{\bf{X}}_{2}=0$.
Thus it is clear that all
$2\times 2$
unitary matrices, which are a complex scalar times that of the 
elements  of the group SU(2), are homogeneous
polarization elements.
These are the so called \emph {rotators} which is obviously so 
because of the known homomorphism between
the group SU(2) and the subgroup 
$1\oplus {\bf{R}}_{3}$ of the group SO(3,1) 
 with $\bf{R}_{3}\in$ SO(3).

Similarly we consider 
another important set of elements belonging to
SL(2,C).
These are the $2\times 2$ 
matrices (of unit determinant) which are hermitian and are 
represented in terms of a (unit) quaternion 
${\bf q}=(q_0,\;\vec{\bf q})$ as 
\begin{equation}  
{\bf{T}}_{b}=\ba{rr} q_{0}-iq_{3} & -iq_{1}-q_{2} \\
                                          -iq_{1}+q_{2} & q_{0}+iq_{3} \ea
\end{equation}
with
$q_{0}$ real and $q_{1}$, $q_{2}$,
$q_{3}$ being purely imaginary.  
The eigenvectors of this matrix are found to be 
$\lambda=q_0\pm \sqrt{q_0^2-1}$ and owing to the 
relations \cite{KNS} $q_0=\cosh {{\theta_b}\over 2}$, 
$q_i=n_i \sinh {{\theta}_b \over 2}$, $i=1,\;2,\;3$ we have 
$\lambda_{1,\;2}=\exp {\pm {\theta_b\over 2}}$. 
The eigenvalues of this matrix are given by 
\begin{equation}
{\bf X}'_1=\ba{c} 1\\ \frac{n_1+in_2}{n_3+1} \ea;\quad 
{\bf X}'_2=\ba{c} 1\\ \frac{n_1+in_2}{n_3-1}\ea
\end{equation}
where $\hat{\bf n}=\{n_1,\;n_2,\;n_3\}$ is a real 
unit vector and ${\bf X}'_1$, ${\bf X}'_2$ are 
mutually orthogonal. We thus have no hesitation in concluding that 
all hermitian $2\times 2$ matrices are 
homogeneous. The angle $\theta_b$ is called the boost angle 
because of the homomorphism that exists between hermitian elements of 
SL(2,C) and the set of all boost matrices 
belonging to the 
group SO(3,1). The elements of the group 
SO(3,1) corresponding to the hermitian 
elements of the group SL(2,C) being given by 
\begin{small}
\begin{equation}{\bf M}_b={\bf A}({\bf T}_b\otimes {\bf T}_b^{*})
{\bf A}^{-1}; \quad {\bf A}=\ba{ccrr} 1 & 0 & 0 & 1 \\
                                      1 & 0 & 0 & -1 \\ 
                                      0 & 1 & 1 & 0 \\
				      0 & i & -i & 0 \\ \ea,
\end{equation}
\end{small}
the Stokes vectors corresponding to the eigenvectors ${\bf X}'_1$ 
and ${\bf X}'_2$ 
are respectively given by    
\begin{equation} {\bf S}'_1={\bf A}({\bf X}'_1\otimes {\bf X}_1^{'*})
=\{1,\;n_3,\;n_1,\;n_2\},  
\quad {\bf S}'_2={\bf A}({\bf X}'_2\otimes {\bf X}_2^{'*})
=\{1,\;-n_3,\;-n_1,\;-n_2\}.  
\end{equation}
${\bf S}'_1$ and ${\bf S}'_2$ being  
the eigenvectors of ${\bf M}_b$ corresponding to the 
eigenvalues $\exp \theta_b$ and $\exp (-\theta_b)$ respectively, the 
reason why these matrices are called diattenuators is obvious. 
They transmit the 
orthogonal vectors (orthogonal in the usual sense, not in the 
Minkowski sense) ${\bf S}'_1$ and ${\bf S}'_2$ with different 
amounts of absorption 
($\exp \theta_b$ for ${\bf S}'_1$ and $\exp (-\theta_b)$ for 
${\bf S}'_2$). 
For the sake of completeness, we write down 
the Stokes vectors ${\bf S}_1$ and ${\bf S}_2$ corresponding to 
the orthogonal eigenpolarizations ${\bf X}_1$, ${\bf X}_2$ of 
${\bf T}_r$. They are, 
\be 
{\bf S}_1=\{1,\;-a_3,\;-a_1,\;-a_2\}\quad \mbox{and}\quad  
{\bf S}_2=\{1,\;a_3,\;a_1,\;a_2\},
\ee
and it is to be noted that whereas  
${\bf X}_1\in \exp \theta/2$ and 
${\bf X}_2\in \exp (-\theta/2)$, the Stokes vectors 
${\bf S}_1$ and 
${\bf S}_2$ belong to the 
eigenvalues (doubly repeated) $1,\;1$ of ${\bf M}_r$.   
The remaining two eigenvectors  
${\bf S}_3$ and ${\bf S}_4$ of ${\bf M}_r$ belonging to the 
eigenvalues $\exp i\theta$, $\exp (-i\theta)$ can be seen to be 
complex 4-vectors thus not corresponding to physical 
light beams. 
Similarly the eigenvectors ${\bf S}'_3$ and ${\bf S}'_4$ belonging 
to the doubly repeated eigenvalues $1,\;1$ of the boost matrix 
do not correspond to physical light beams as both of them are 
not Minkowskian vectors. 

Having thus arrived at the conclusions that rotation and boost 
matrices are the homogeneous polarization elements having 
orthogonal eigenvectors, we now wish to see which other 
elements of the Lorentz group correspond to homogeneous 
polarization elements. 
To make this examination, we find it useful to recall the 
well-known polar decomposition theorem \cite{LT} realised 
in the case of the group SL(2,C). 
We notice that any element  
$\bf T$ of the group SL(2,C) can be written as a product of a 
boost matrix ${\bf T}_b$ (or ${\bf T}'_b$)and a rotation matrix ${\bf T}_r$ as shown 
below. 
\begin{equation}{\bf T}={\bf T}_r{\bf T}_b={\bf T}'_b{\bf T}_r
\end{equation}
The eigenvectors belonging to ${\bf T}_r$ and ${\bf T}_b$ 
respectively being given by equation (2.2) and (2.4), it is not 
difficult to see that for $\bf T$ to possess orthogonal eigenvectors, 
one should have 
$\hat{\bf a}=\hat{\bf n}$. 
Also, by using the quaternionic representation of the Lorentz 
group, one can very easily get at the general form of the 
polarization element possessing orthogonal eigenpolarizations. 
Since the quaternions corresponding to ${\bf T}_r$, 
${\bf T}_b$ are, respectively, 
${\bf q}_r=\left( \cos {\theta_{r} \over 2},
\;\hat{\bf a}\sin {\theta_{r} 
\over 2} \right)$ and 
${\bf q}_b=\left( \cosh {\theta_b \over 2},\;i\hat{\bf n}\sinh 
{\theta_b\over 2}\right )$, by using the rule of multiplication 
of quaternions and using the condition $\hat{\bf a}=\hat{\bf n}$, 
we get at the quaternion ${\bf q}$  
corresponding to $\bf T$, which is homogeneous. It is given by 
\begin{eqnarray}
{\bf q}={\bf q}_r{\bf q}_b  ;\quad 
{\bf q}=\left( q_0,\;\vec{\bf q}\right);\nonumber \\
q_0=\cos {\theta_r \over 2} \cosh {\theta_b \over 2}-
i\sin {\theta_r \over 2} \sinh {\theta_b \over 2}, \\
\vec{\bf q}=\left( \cos {\theta_r \over 2} 
\sinh {\theta_b \over 2}+
\sin {\theta_r \over 2} \cosh {\theta_b \over 2}\right)
\hat{\bf a}.\nonumber
\end{eqnarray}
One can very easily see that when $\theta_r=0$, 
${\bf q}$ given above 
reduces to ${\bf q}_b$, the quaternion corresponding to a boost
and when $\theta_b=0$, $\bf q$ reduces to 
${\bf q}_r$, the quaternion corresponding to a rotation. 
When both $\theta_r$ and $\theta_b$ are non-zero, 
the quaternion which obeys (2.9) is the one 
corresponding to the general 
homogeneous polarization element. 
At this stage, an observation regarding the classification of the 
group SO(3,1) depending on the geometic structure its elements possess, 
may be in order. 
We recall that \cite{KNS} if an element 
${\bf L}=\exp ({\bf S})$ of SO(3,1), where $\bf S$ is the
so-called \emph{infinitisimal transformation matrix} of 
${\bf L}$, has an additional structure 
with 
\[
{\bf S}\equiv {\bf Y}\widetilde{\bf X}-
{\bf X}\widetilde{\bf Y};\quad \widetilde{\bf X}{\bf Y}=0, 
\]
${\bf X}$, ${\bf Y}$ being Minkowski 4-vectors,   
then it is 
called a \emph{planar} Lorentz transformation. All other elements 
of SO(3,1) which do not have their corresponding $\bf S$ in the 
form given above are called \emph{non-planar} Lorentz 
transformations. 
A corresponding classification of the group SL(2,C), though  
this classification is significant mostly  
in the $4\times 4$ representation of the Lorentz group, is obvious. 
Also, depending on the 4-vector character of the 
vectors $\bf X$ and $\bf Y$, we have the so-called \cite{KNS}
\emph{rotation-like}, \emph{null} and \emph{boost-like} planar 
transformations. The quaternions corresponding to planar 
and 
non-planar Lorentz transformations have been identified \cite{KNS} 
and it is seen that `rotations' and `boosts' which are identified 
to be homogeneous 
polarization elements are respectively special cases of 
`rotation-like' and `boost-like' Lorentz transformations \cite{KNS}. 
But it is 
interesting to note that \textbf{none of the other planar 
transformations are 
homogeneous}. This can be seen by observing that 
the quaternions corresponding to planar Lorentz transformations 
have $q_0$ real as the only general 
condition on them, with 
their other quaternion components $q_1$, $q_2$ and $q_3$ being 
permitted to take any values subject to the condition 
${\bf q}\cdot {\bf q}=1$.        
A careful observation of 
equation (2.9) reveals that the quaternions corresponding 
to planar Lorentz
transformations do not coincide with the form of the 
quaternion given in that equation (equation (2.9)) for any value of $\theta_r$ and 
$\theta_b$ except when either $\theta_r=0$ or $\theta_b=0$. 
The situations in equation (2.9) when $\theta_b=0$ corresponding 
to the quaternion representing a rotation and the other 
situation when $\theta_r=0$ corresponding to the quaternion 
representing a boost, our assertion made above is proved. 
Thus, with the set of all homogeneous elements in the Lorentz 
group being identified by equation (2.9), it containing planar Lorentz 
transformations only in the form of rotations and boosts, the 
remaining elements corresponding to non-planar Lorentz transformations, 
we can conclude that planar Lorentz transformations with the 
exception of  
rotations and boosts are either inhomgeneous or degenerate 
polarization elements.  
In the following we try to identify the set of all 
degenerate polarization elements in the Lorentz group. 

\subsection{Non-singular Degenerate Polarization elements}
We start with checking the so called \emph{null} 
elements \cite{KNS} 
of the Lorentz group for their \emph{`degenerate'}ness. 
The null elements of the Lorentz group are the ones which have 
their corresponding quaternion as 
\be
{\bf q}_n=\left( q_0,\;\vec{\bf q} \right);\quad 
q_1^2+q_2^2+q_3^2+q_0^2=1,
		  \quad q_1^2+q_2^2+q_3^2=0.
\ee  
One can easily check that for all elements of the Lorentz group 
having the quaternion of the above form,  
the number $1$ appears as the doubly degenerate eigenvalue. The 
eigenvector belonging 
to this doubly repeated eigenvalue can be seen to be 
\be {\bf X}_0=\ba{c} 1 \\ {-iq_1 \over iq_2+q_3} \ea,
\ee 
thus taking the null elements ${\bf T}_n$ into the class of 
degenerate polarization elements. 
None of the other elements 
of the Lorentz group are degenerate as evidenced by the fact that 
only the null elements 
of the group have doubly 
degenerate eigenvalues. 
This result and the discussions made hitherto  
make it clear that the `null' elements of the Lorentz group are the
only non-singular degenerate polarization elements (apart from a 
scale factor).
Having thus identified the \emph {disjoint} sets of the Lorentz 
group corresponding to homogeneous and degenerate 
polarization elements, we can now conclude that all other 
elements of the Lorentz group belong to the class of inhomogeneous 
polarization elements. Or to put it more simply, the elements of the 
Lorentz group which 
have the corresponding quaternions other than the ones mentioned 
in equations (2.9) and (2.10), correspond to inhomogeneous 
polarization elements.   

It is worthwhile to point out here that all the 
examples of inhomogeneous and degenerate polarization elements 
that are quoted in \cite{LC} are singular elements though there is 
a whole lot of non-singular inhomogeneous as well as degenerate 
polarization elements as we have pointed out here.      
In Table~1 we give a few illustrative examples of homogeneous, 
inhomogeneous and degenerate polarization elements which are 
non-singular. 

\section {Singular Mueller matrices}
\setcounter{equation}{0}
Having till now studied the eigenstructure and hence the 
classification of non-singular polarization elements, we now seek 
to determine the eigenstructure of singular polarization elements. 
It is to be noted here that a class of the widely studied and used 
polarizing optical devices, the so called 
\emph {polarizers and analyzers}\footnote{For a discussion on 
polarizers and analyzers, see \cite{Thesis, LC2}.} 
are singular. Thus it is worthwhile to study their eigenstructure 
and eventually classify them on that basis. 
Here we wish to work in the $4\times 4$ representation of the 
devices only as it is convenient mathematically for the task 
we have on hand. 

We recall that \emph{singular pure Mueller matrices} have the form 
\begin{eqnarray} {\bf M}=m_{00}{\bf X}\widetilde{\bf Y}{\bf G}, 
\quad \mbox {where}\quad {\bf X}=\{1,\;x,\;y,\;z\}; \;
\widetilde{\bf X}{\bf G}{\bf X}=0, \nonumber \\
 {\bf Y}=\{1,\;-p,\;-q,\;-r\}; \;
\widetilde{\bf Y}{\bf G}{\bf Y}=0,\; \mbox{and}\; 
{\bf G}=\mbox{diag}\left(1,\;-1,\;-1,\;-1\right).
\end{eqnarray}
It can be seen that matrices of the form ${\bf M}=m_{00}{\bf X}\widetilde{\bf Y}{\bf G}$ 
possess the following eigenvalues 
\begin{equation}
\lambda_1=m_{00}(1+px+qy+rz),\; \lambda_2=0,0,0.
\end{equation}
Thus, there are only two distinct eigenvalues for $\bf M$ of the form (3.1)
in general and a quadruply repeated eigenvalue $\lambda=0$ in the case where 
${\bf X}={\bf Y}$.  Correspondingly, in general, there are two eigenvectors 
${\bf S}_1={\bf X}\in \lambda_1$ and 
${\bf S}_2={\bf Y}\in \lambda_2$, the case of there being only 
one eigenvector $\bf Y\in \lambda=0$ getting realized when 
${\bf X}={\bf Y}$. 
Since 
$\widetilde {\bf S}_1{\bf S}_2=\widetilde{\bf X}{\bf Y}$ 
is the analogue of the expression  
${\bf X}_1^\dagger{\bf X}_2$, where 
${\bf X}_1$ and 
${\bf X}_2$ are the eigenvectors of the Jones matrix 
corresponding to ${\bf M}$ in equation (3.1), singular pure Mueller 
matrices fall into the class of homogeneous, inhomogeneous 
or degerate 
polarization elements depending on the vectors 
${\bf X}={\bf S}_1$ and ${\bf Y}={\bf S}_2$ on which they are built.  
They are homogeneous when $\widetilde{\bf X}{\bf Y}=0$, inhomogeneous when 
$\widetilde{\bf X}{\bf Y}\neq 0$ and degenerate when ${\bf X}={\bf Y}$.
In Table~2, we give few examples of homogeneous, inhomogeneous and 
degenerate polarization elements which are singular. One can also 
see \cite{LC} for several examples of singular inhomogeneous 
polarization elements. 

Though we do not have corresponding Jones matrices for 
singular Mueller matrices other than the ones mentioned in 
equation (3.1), we find it worthwhile to examine the 
eigenstructure of 
a special class ${\cal M}$ of singular Mueller matrices 
which have the same structure as that of the singular pure 
Mueller matrices, but the 4-vector character of the composite vectors 
$\bf X$ and ${\bf Y}$ are different from that of singular pure 
Mueller matrices. We have three cases to consider depending on the choices 
possible for the 4-vectors ${\bf X}$ and $\bf Y$. 
\begin{enumerate}
\item[(i)]
Here ${\bf M}_1\in {\cal M}$ is of the form 
\begin{eqnarray} {\bf M}_1=m_{00}{\bf X}\widetilde{\bf Y}{\bf G}, 
\quad \mbox {where}\quad \widetilde{\bf X}{\bf G}{\bf X}=0, 
\nonumber  \\
\widetilde{\bf Y}{\bf G}{\bf Y}>0,\; \mbox{and} \;
{\bf G}=\mbox{diag}\left(1,\;-1,\;-1,\;-1\right).
\end{eqnarray}
It is easy to see that this matrix has only one non-zero 
eigenvalue $\lambda_1=m_{00}(1+px+qy+rz)$
its corresponding eigenvector being ${\bf S}_1={\bf X}$. 
The eigenvectors corresponding to its triply repeated zero 
eigenvalue $\lambda_2=0$ can be seen to be non-Stokes. 
It may be of some interest to note that this 
matrix is the so called \emph {generalized polarizer} matrix 
\cite{Thesis}. 
\item[(ii)]  Consider a $4\times 4$ matrix ${\bf M}_2$  where 
\begin{eqnarray} 
{\bf M}_2=m_{00}{\bf X}\widetilde{\bf Y}{\bf G}; 
\quad \widetilde{\bf X}{\bf G}{\bf X}>0, 
\nonumber  \\
\widetilde{\bf Y}{\bf G}{\bf Y}=0,\; \mbox{and} \;
{\bf G}=\mbox{diag}\left(1,\;-1,\;-1,\;-1\right).
\end{eqnarray}  
This matrix also possesses only one non-zero eigenvalue 
It is easy to notice that ${\bf M}_2$ has a structure equivalent to 
that of the transpose of ${\bf M}_1$.
It has got two eigenvalues $\lambda_1\neq 0$ and $\lambda_2=0$ 
and the corresponding eigenvectors are 
${\bf S}_1={\bf X}\in \lambda_1$, ${\bf S}_2={\bf Y}\in \lambda_2=0$. 
The form of the above matrix itself suggests that it belongs to the 
so called \emph{generalized analyzer matrix} \cite{Thesis}. 

\item[(iii)] The only other remaining possibility in the 
choice of $\bf X$ and $\bf Y$ being 
$\widetilde{\bf X}{\bf G}{\bf X}>0$ and 
$\widetilde{\bf Y}{\bf G}{\bf Y}>0$, we have 
\begin{eqnarray} {\bf M}_3=m_{00}{\bf X}\widetilde{\bf Y}{\bf G}; 
\quad \widetilde{\bf X}{\bf G}{\bf X}>0, 
\nonumber  \\
\widetilde{\bf Y}{\bf G}{\bf Y}>0,\; \mbox{and} \;
{\bf G}=\mbox{diag}\left(1,\;-1,\;-1,\;-1\right).
\end{eqnarray} 
On the same lines of that of the previous two cases, we can see that 
the matrix ${\bf M}_3$ has two distinct eigenvalues, a non-zero eigenvalue 
$\lambda_1$ and a zero eigenvalue $\lambda_2=0$.  
The only eigenvector corresponding to $\lambda_1$ 
is ${\bf X}$ whereas the eigenvectors belonging to the triply repeated eigenvalue 
$\lambda_2=0$ are all 
seen to be non-Stokes vectors and thus not qualifying to be called eigenpolarizations.  
\end{enumerate} 

We wish to make an observation here on the importance of 
studying the eigenstructre of Mueller matrices. The 
eigenvector which correspond to a real eigenvalue of 
a given Mueller matrix represents the physical light beam that  
comes out undisturbed by the polarizing optical device represented 
by the Mueller matrix and hence a study of the eigenstructure of 
Mueller matrices, at least in the cases possible, is welcome 
for understanding the nature of the optical devices represented 
by them. In fact, the singular Mueller matrices 
that we have studied here is one class of Mueller matrices whose 
eigenstructure can be studied quite easily. But there still remain a 
whole lot of Mueller matrices, singular as well as non-singular, 
which remain to be examined for their eigenstructure. Among the 
class of non-singular Mueller matrices, we have carried out a 
study of the pure Mueller matrices which are elements of the 
Lorentz group (apart from a scale factor) and are in pursuit of 
other non-singular  
Mueller matrices which are accessible for a study of their 
eigenstructure.

\noindent{\bf Acknowledgements }
One of the authors, Dr. Sudha thanks the CSIR for the 
financial support in the form of Research Fellowship.

\begin{table}
\caption{Non-singular Polarization Elements}
\begin{center}
\begin{tabular}{||c|c|c||}
\hline\hline
Type of & & \\ 
polarization & Example 1 & Example 2 \\
element & & \\
\hline
& & \\
 & $\frac{1}{\sqrt{2}}\ba{rr}\sqrt{2}-i & 1- \sqrt {2}i \\
1- \sqrt {2}i & \sqrt{2}-i \ea $  & $ \frac{1}{2}\ba{rr} 
2-3i & -2\sqrt {3}-\sqrt {3}i \\ 
2\sqrt {3}+\sqrt {3}i & 2-3i \ea $   \\
Homogeneous & & \\
&  ${\bf X}_1=\frac{1}{\sqrt{2}}\{1,\;-1\}$; 
${\bf X}_2=\frac{1}{\sqrt{2}}\{1,\;1\}$ & 
${\bf X}_1=\frac{1}{\sqrt{2}}\{1,\;-i\}$; 
${\bf X}_2=\frac{1}{\sqrt{2}}\{1,\;i\}$ \\
& & \\
\hline
& & \\
 & $\ba{rr}-i & -1- 2i \\
1 & 2+i \ea $  & $\ba{rl} 2+\sqrt{3} & 0 \\
2(1-i) & 2-\sqrt{3} \ea $  \\
Inhomogeneous & & \\
&  ${\bf X}_1=\frac{1}{\sqrt{2}}\{1,\;-1\}$; 
${\bf X}_2=\frac{\sqrt{5}}{\sqrt{6}}\{1,\;\frac{-1}{1+2i}\}$ & 
${\bf X}_1=\frac{\sqrt{3}}{\sqrt{5}}\{1,\;\frac{1-i}{\sqrt{3}}\}$; 
${\bf X}_2=\{0,\;1\}$ \\ 
& & \\
\hline \\
& & \\
 &$\ba{rl}-i & 1- i \\
1-i & 2+i \ea $  & $\ba{rl} 1 & 0 \\
2(1-i) & 1 \ea $  \\
Degenerate & & \\
&  ${\bf X}_1={\bf X}_2=\frac{1}{\sqrt{2}}\{1,\;i\}$;  & 
${\bf X}_1={\bf X}_2=\{0,\;1\}$; \\
& & \\
\hline\hline
\end{tabular}
\end{center}
\end{table}

\begin{table}
\caption{Singular Polarization Elements}
\begin{center}
\begin{tabular}{||c|c|c||}
\hline\hline
Type of & & \\ 
polarization & Example 1 & Example 2 \\
element & & \\
\hline
& & \\
 & $\ba{rr} 1 & 1 \\
1 & 1 \ea $  & $ \ba{rr} 
0 & 0 \\ 
0 & 1 \ea $   \\
Homogeneous & & \\
&  ${\bf X}_1=\frac{1}{\sqrt{2}}\{1,\;1\}$; 
${\bf X}_2=\frac{1}{\sqrt{2}}\{1,\;-1\}$ & 
${\bf X}_1=\{0,\;1\}$; 
${\bf X}_2=\{1,\;0\}$ \\
& & \\
\hline
& & \\
 & $\ba{rr} 1 & i \\
1 & i \ea $  & $\ba{rl} 1 & 0 \\
1 & 0 \ea $  \\
Inhomogeneous & & \\
&  ${\bf X}_1=\frac{1}{\sqrt{2}}\{1,\;1\}$; 
${\bf X}_2=\frac{1}{\sqrt{2}}\{1,\;i\}$ & 
${\bf X}_1=\frac{1}{\sqrt{2}}\{1,\;1\}$; 
${\bf X}_2=\{0,\;1\}$ \\ 
& & \\
\hline \\
& & \\
 & $\ba{rl}1 & -1 \\
1 & -1 \ea $  & $\ba{rl} 0 & 1 \\
0 & 0 \ea $  \\
Degenerate & & \\
&  ${\bf X}_1={\bf X}_2=\frac{1}{\sqrt{2}}\{1,\;1\}$;  & 
${\bf X}_1={\bf X}_2=\{1,\;0\}$; \\
& & \\
\hline\hline
\end{tabular}
\end{center}
\end{table}

\end{document}